\documentclass[letterpaper,preprintnumbers,prd,twocolumn,nofootinbib,nobibnotes,showpacs]{revtex4}
\usepackage{amsfonts}
\usepackage{mathrsfs}
\usepackage{epsfig}
\usepackage{graphicx}%
\usepackage{dcolumn}
\usepackage{amsmath}
\usepackage{color}
\usepackage{color}
\makeatletter
\def\btt#1{\texttt{\@backslashchar#1}}%
\DeclareRobustCommand\bblash{\btt{\@backslashchar}}%
\makeatother
\begin{document}

\title{Cosmic evolution of scalar fields with multiple vacua: generalized DBI and quintessence}
\author{Changjun Gao}\email{gaocj@bao.ac.cn} \affiliation{Key Laboratory for Computational Astrophysics, National Astronomical
Observatories, Chinese Academy of Sciences, Beijing, 100012,
China} \affiliation{State Key Laboratory of Theoretical Physics,
Institute of Theoretical Physics, Chinese Academy of Sciences,
Beijing 100190, China}
\author{You-Gen Shen}
\email{ygshen@center.shao.ac.cn} \affiliation{Shanghai
Astronomical Observatory, Chinese Academy of Sciences, Shanghai
200030, China}

\date{\today}

\begin{abstract}
We find a method to rewrite the equations of motion of scalar
fields, generalized DBI field and quintessence, in the autonomous
form for\emph{arbitrary} scalar potentials. With the aid of this
method, we explore the cosmic evolution of generalized DBI field
and quintessence with the potential of multiple vacua. Then we
find that the scalars are always frozen in the false or true
vacuum in the end. Compared to the evolution of quintessence, the
generalized DBI field has more times of oscillations around the
vacuum of the potential. The reason for this point is that, with
the increasing of speed $\dot{\phi}$, the friction term of
generalized DBI field is greatly decreased. Thus the generalized
DBI field acquires more times of oscillations.

\end{abstract}


\maketitle

\section{Introduction}
Scalar fields play an important role in both theoretical physics
and modern cosmology for their simple but non-trivial dynamics. In
theoretical physics, they are present in the Jordan-Brans-Dicke
theory as Jordan-Brans-Dicke scalar \cite{brans:61}; in
Kaluza-Klein compactification theory as the radion
\cite{csaki:00}; in the Standard Model of particle physics as the
Higgs boson \cite{higgs:64}; in the low-energy limit of the
superstring theory as the dilaton \cite{gib:88}, tachyon
\cite{sen:02} and DBI (Dirac-Born-Infeld) field \cite{ward:07}. In
cosmology, scalar fields are employed to model the inflaton
\cite{guth:81}, the quintessence
\cite{fujii:88,ratra:88,chiba:97,fer:97,cope:98,cald:98,zla:99}
(for a recent review of\emph{ quintessence}, see Refs.
\cite{tsu:13} and references therein), the k-essence
\cite{chiba:00,arm:00}, the phantom \cite{cald:02}, in order to
drive the inflation of the early Universe or to speed up the
expansion of the late Universe. We note that most of the scalar fields are
 hypothetical particles. But the Higgs Boson has uniquely been discovered by experiments.

Quintessence is a canonical scalar field which is assumed to be
minimally coupled to gravity. Compared to other scalar fields,
quintessence turns out to be the simplest scenario which is free
of ghosts and instability problems. The dynamics of quintessence
in the presence of matters has been studied in great detail for
many different potentials
\cite{cope:98,cald:98,zla:99,mac:00,ng:01,cor:03,cald:05,linder:06,
bar:00,cope:09,liddle:99,sahni:00IJ,sahni:00PRD}. However, for a
general quintessence potential, the equations of motion are rather
involved. To our knowledge, one have not yet find a general method
to write the equations of motion in the autonomous form for
\emph{arbitrary} potential. Thus the purpose of this article is to
report that we have found a way.

The DBI field describes the dynamics of D-branes evolving in a
higher-dimensional warped spacetime. A novel aspect of this field
is the existence of a speed limit on the field space, resulting
from causality restrictions on the motion of the branes in the
bulk spacetime. The speed limit is enforced by the non-canonical
kinetic terms in the DBI field. This is different from the
quintessence whose speed $\nabla{\phi}$ could be arbitrarily
large. From this point of view, quintessence and DBI field are the
counterpart of Newtonian and Special Relativity mechanics,
respectively. The cosmic evolution of DBI field have been studied
in Refs.~\cite {guo:08}. These researches only apply to some
special forms of potentials. Thus, to find a general method
applying to \emph{arbitrary} DBI potential constitutes the second
purpose of this article.

\section{generalized DBI field}
We consider a $\textrm{D}3$-brane with tension ${\textit{T}}$
evolving in a 5-dimensional spacetime. The dynamics of the mobile
D3-brane is described by the DBI action. The D3-brane is free to
move on the internal compact Calabi-Yau manifold. The generalized
DBI action can be written as follows \cite{ward:07}
\begin{eqnarray}\label{eq:action}
S&=&\int
d^4x\sqrt{-g}\left[T\left(\psi\right)W\left(\psi\right)\sqrt{1+\frac{1}{T\left(\psi\right)}\partial_{\mu}\psi\partial^{\mu}\psi}
\right.\nonumber\\&&\left.-T\left(\psi\right)+{V}\left(\psi\right)\right]+S_{m}\;.
\end{eqnarray}
Here $T(\psi)$ is the warped tension of the brane and $S_m$ is the
action for matters localized in the D3-brane. The potential
$W(\psi)$ could arise under the condition that the brane is a
non-BPS one \cite{sen:02} or there are multiple coincident branes
\cite{myers:99}. The potential ${V}(\psi)$ is related to the
brane's interactions with bulk fields or other branes.

In order to simplify our derivations, we define the variable
$\phi$ as follows
\begin{eqnarray}
\frac{1}{\sqrt{T\left(\psi\right)}}\partial_{\mu}\psi=\partial_{\mu}\phi\;.
\end{eqnarray}
Then above action can be written as
\begin{eqnarray}\label{eq:action1}
S&=&\int
d^4x\sqrt{-g}\left[T\left(\phi\right)W\left(\phi\right)\sqrt{1+\partial_{\mu}\phi\partial^{\mu}\phi}
\right.\nonumber\\&&\left.-T\left(\phi\right)+{V}\left(\phi\right)\right]+S_{m}\;.
\end{eqnarray}
Without the loss of physical significance, we could absorb the
term $T(\phi)$ into $W(\phi)$ and $V(\phi)$, respectively. Then we
find the action is simply

\begin{eqnarray}\label{eq:action2}
S&=&\int
d^4x\sqrt{-g}\left[W\left(\phi\right)\sqrt{1+\partial_{\mu}\phi\partial^{\mu}\phi}
+{V}\left(\phi\right)\right]+S_{m}\;.
\end{eqnarray}

We shall investigate the cosmic evolution of the DBI field in the
background of spatially flat Friedmann-Robertson-Walker Universe
\begin{eqnarray}
ds^2=-dt^2+a\left(t\right)^2\left(dr^2+r^2d\Omega^2\right)\;,
\end{eqnarray}
where $a(t)$ is the cosmic scale factor. We model the matter
sources present in the Universe as perfect fluids. The perfect
fluids can be baryonic matter, relativistic matter and dark
energy. We assume there is no interaction between the generalized
DBI scalar field and the matter fields, other than by gravity.
Then the Einstein equations and the equation of motion of the
scalar field are given
\begin{eqnarray}\label{eq:ein00}
&&3H^2=\kappa^2\left(W\gamma +V+\rho_m\right)\;,\nonumber\\
&&2\dot{H}+3H^2=-\kappa^2\left(-W/\gamma-V+\omega_m\rho_m\right)\;,
\end{eqnarray}
and
\begin{eqnarray}\label{eq:ein11}
\ddot{\phi}+\frac{1}{\gamma^2}\cdot3H\dot{\phi}+\frac{1}{\gamma^3}\cdot\frac{V_{,\phi}}{W}+\frac{1}{\gamma^2}\cdot\frac{W_{,\phi}}{W}=0\;,
\end{eqnarray}
respectively. Here $H\equiv\dot{a}/a$ denotes the Hubble parameter
and dot is the derivative with respect to cosmic time, $t$.
$\rho_m$ and $\omega_{m}$ are the energy density and the equation
of state for the matter sources. We have $\omega_m=-1,\ 0,\ 1/3,\
+1$ for the cosmological constant, dust matter, relativistic
matter and stiff matter, respectively. In this paper, we shall
consider the case of dust matter, $\omega_m=0$. $V_{,\phi}$ and
$W_{,\phi}$ denote the derivative with respect to $\phi$. $\gamma$
is defined by
\begin{eqnarray}
\gamma=\frac{1}{\sqrt{1-\dot{\phi}^2}}\;,
\end{eqnarray}
which has the physical meaning of the generalized Lorentz factor.
It is apparent the speed of scalar $\dot{\phi}$ is constrained to
be smaller than the speed of light. This is remarkably different
from the usual quintessence which could have arbitrary large speed
in the sense of classical mechanics.

Observing Eqs.~(\ref{eq:ein00}) and Eq.~(\ref{eq:ein11}), we could
absorb the constant $\kappa^2/3$ ($\kappa^2=8\pi$) into $W, V$ and
$\rho_m$, respectively,

\begin{eqnarray}
&&W\longrightarrow W\cdot\frac{3}{\kappa^2}\;,\\
&&V\longrightarrow V\cdot\frac{3}{\kappa^2}\;,\\
&&\rho_m\longrightarrow \rho_m\cdot\frac{3}{\kappa^2}\;.
\end{eqnarray}
Then above equations of motion turns out to be
\begin{eqnarray}\label{eq:ein0}
&&H^2=W\gamma +V+\rho_m\;,\\
&&2\dot{H}/3+H^2=W/\gamma+V-\omega_m\rho_m\;,\\
&&\ddot{\phi}+\frac{1}{\gamma^2}\cdot3H\dot{\phi}+\frac{1}{\gamma^3}\cdot\frac{V_{,\phi}}{W}+\frac{1}{\gamma^2}\cdot\frac{W_{,\phi}}{W}=0\;.
\end{eqnarray}
Given the scalar potential $W(\phi)$, $V(\phi)$ and the equation
of state $\omega_m$, we are left with three variables,
$a(t),\rho_m$ and the DBI field $\phi(t)$. Then we have three
variables and three independent differential equations. So the
system of equations is closed.

It is rather difficult to find the analytic solutions to the
equations of motion (12-14). Hence in order to solve them
numerically, we had better rewrite them in the autonomous form. To
this end, we introduce the following dimensionless quantities
\begin{eqnarray}
&&X\equiv\dot{\phi}\;,\ \ \ \ \ \ \ \ \  U \equiv\frac{{V_{,\phi}}}{W^{\frac{3}{2}}}\;,\\
&&Y\equiv\frac{\sqrt{W}}{H}\;,\ \ \ \ \ Q \equiv\frac{{W_{,\phi}}}{W^{\frac{3}{2}}}\;,\\
&&Z\equiv\frac{\sqrt{V}}{H}\;,\ \ \ \ \ \ \ S
\equiv\frac{{V_{,\phi}}}{V^{\frac{3}{2}}}\;,\\
&&N\equiv\ln a\;.
\end{eqnarray}
We see $U,\ Q,\ S$ are the functions of DBI field, $\phi$. So they
can be expressed as the function of $Y/Z$. Now we have only three
variables, namely, $X,\ Y,\ Z$ and it is sufficient for us to
derive the corresponding three independent differential equations.
We remember that the system of equations of $X,\ Y\, Z$ are an
autonomous system if and only if $U,\ Q,\ S$ are expressed as the
functions of $X,\ Y,\ Z$. For simplicity, we assume
\begin{eqnarray}
\frac{V}{W}=\phi^2\;.
\end{eqnarray}
We emphasize that one could in general assume ${V}/{W}=F(\phi)$
with $F(\phi)$ other simple functions, for example, $F(\phi)=\phi^{n}, \ e^{\alpha\phi},\ \ln \phi$ and so on. Then we could
obtain $\phi=\phi\ (V/W)=\phi\ (Z/Y)$ from this equation.
Similarly, $U,\ Q,\ S$ could be expressed as the function of
$Z/Y$.

Now with the aid of this assumption, we are able to write the
Eqs.~(12-14) in the autonomous form and  deal with \emph{any}
scalar potential, $V(\phi)$ in the calculations. In this article,
we shall focus on the scalar potential $V(\phi)$ with the
expression of
\begin{eqnarray}\label{eq:potential}
V&=&V_0+\frac{V_1}{\phi^6}\left(\phi-a_1\right)\left(\phi-a_2\right)\left(\phi-a_3\right)
\nonumber\\&&\cdot\left(\phi-a_4\right)\left(\phi-a_5\right)\left(\phi-a_6\right)\;,
\end{eqnarray}
where $V_i$ and $a_i$ are all positive constants. The reason for
this choice of potential is that what we want to study is a
potential with multiple vacua. Scalar fields with multiple vacua
are very interesting because they have both theoretical origin in
string landscape \cite{landscape:03} and theoretical study in
Coleman-De Luccia tunnelling \cite{coleman:77}.

We note that the case of the well-known AdS throat,
$W(\phi)=\lambda/\phi^4$ has been included in the desired one. As
an example, we put $V_0=2,\ V_1=10^4,\ a_1=1,\ a_2=11/10,\
a_3=9/8,\ a_4=39/25,\ a_5=157/100,\ a_6=2$ in the following
discussions.

\begin{figure}[h]
\begin{center}
\includegraphics[width=9cm]{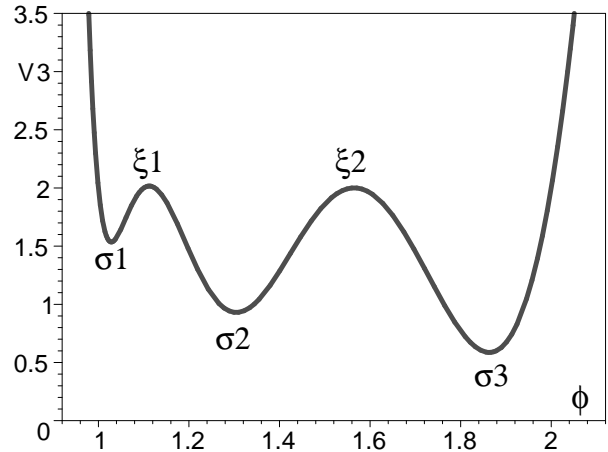}
\caption{The DBI potential $V(\phi)$ with respect to $\phi$. There
are two local maximum ($\xi_1=1.11,\ \xi_2=1.56$), one real vacuum
($\sigma_3=1.86$) and two false vacuum ($\sigma_1=1.03,\
\sigma_2=1.31$) for the potential. Physically, the scalar field
would roll down the potential and then passes through the first
($\sigma_1$) and the second ($\sigma_2$) false vacuum. Finally, it
arrives at the real vacuum ($\sigma_3$).}. \label{f1}
\end{center}
\end{figure}

\begin{table*}[t]
\begin{center}
\begin{tabular}{|c|c|c|c|c|c|c|c|c|}
\hline Name &  $X$ & $Y$ &$Z$ & $\phi$
 & Stability & $\Omega_\phi$
 & $w_\phi$  \\
\hline \hline (a) & $0$ & $0.70$ & $0.72$ & $\sigma_1=1.03$ &
Stable spiral (attractor)
& $1$ & $-1$\\
\hline \hline (b) & $0$ & $0.67$ & $0.74$ & $\xi_1=1.11$ & Saddle
point
& $1$ & $-1$\\
\hline \hline (c) & $0$ & $0.61$ & $0.79$ & $\sigma_2=1.31$ &
Stable spiral (attractor)
& $1$ & $-1$\\
\hline \hline (d) & $0$ & $0.54$ & $0.84$ & $\xi_2=1.56$ & Saddle
point
& $1$ & $-1$\\
\hline \hline (e) & $0$ & $0.47$ & $0.88$ & $\sigma_3=1.86$ &
Stable spiral (attractor)
& $1$ & $-1$\\
\hline
\end{tabular}
\end{center}
\caption[crit]{Properties of the critical points for the
  scalar potential given by
  Eq.~(20).} \label{crit0}
\end{table*}
In Fig.~{\ref{f1}}, we plot the potential $V(\phi)$ with respect
to $\phi$. There are two local maximum ($\xi_1=1.11,\
\xi_2=1.56$), one real vacuum ($\sigma_3=1.863$) and two false
vacuum ($\sigma_1=1.029,\ \sigma_2=1.305$) for the potential.
Physically, the scalar field would roll down the potential and
then passes through the first ($\sigma_1$) and the second
($\sigma_2$) false vacuum. Finally, it arrives at the real vacuum
($\sigma_3$). The detail of the trajectory is closely related to
the initial velocity, $\dot{\phi_i}$ (with the initial value,
$\phi_i$ fixed). When the initial speed $\dot{\phi_i}$ is small
enough, the scalar would acquire damped oscillations due to the
Hubble friction in the first false vacuum and finally it is frozen
in this vacuum. However, with the increasing of initial speed, the
scalar could cross the first local maximum ($\xi_1$) and finally
is frozen in the second false vacuum ($\sigma_2$). With even much
larger initial velocity, the scalar could cross the two local
maximum ($\xi_1$ and $\xi_2$) and finally dwells on the real
vacuum ($\sigma_3$). Since the speed of the scalar is constrained
to be smaller than the speed of light, the scalar can not climb
the hill with arbitrary height. Due to the Hubble friction, we
expect the fate of the scalar is to dwell on the real vacuum. In
what follows, we shall show theses points numerically.

Using the dimensionless variables defined in (15-18), the
equations of motion (12-14) can be written in the following
autonomous form
\begin{eqnarray}
\frac{dX}{dN}&=&-\frac{1}{\gamma^2}\cdot 3X-\frac{1}{\gamma^3}\cdot UY-\frac{1}{\gamma^2}\cdot QY\;,\nonumber\\
\frac{dY}{dN}&=&\frac{1}{2}QXY^2-Y\frac{\dot{H}}{H^2}\;,\nonumber\\
\frac{dZ}{dN}&=&\frac{1}{2}SXZ^2-Z\frac{\dot{H}}{H^2}\;,
\end{eqnarray}
where $U,\ Q,\ S$ are the functions of $Y/Z$ and
\begin{eqnarray}
\gamma&=&\frac{1}{\sqrt{1-X^2}}\;,\\
\frac{\dot{H}}{H^2}&=&3Y^2\cdot\frac{1-\gamma^2}{2\gamma}-\frac{3}{2}\left(1-\gamma
Y^2-Z^2\right)\;.
\end{eqnarray}
The Friedmann equation becomes the constraint equation
\begin{eqnarray}
1=\frac{Y^2}{\sqrt{1-X^2}}+Z^2+\frac{\rho_m}{H^2}\;.
\end{eqnarray}
The equation of state $w$ of the DBI scalar field is
\begin{eqnarray}
w&\equiv&\frac{-1/\gamma-Z^2/Y^2}{\gamma+Z^2/Y^2}\;.
\end{eqnarray}
In Table I, we present the properties of the five fixed points for
the scalar field. The points (a,\ c,\ e) correspond to the false
vacua ($\sigma_1,\ \sigma_2$) and real vacuum ($\sigma_3$). The
three points are stable spirals.  On these epoches, the scalar
field behaves as a damping oscillator with the equation of state
of firstly behaving as the dust matter and then oscillating
approaching $-1$ (see Fig.~(5)). The points (b,\ d,) correspond to
the two local maximum ($\xi_1,\ \xi_2$) and they are saddle
points. On these points, the speed of the scalar field exactly
vanishes and the DBI field acquires the equation of state of
cosmological constant.

In Fig.~(2-4), we plot the evolution of the speed, $\dot{\phi}$ of
the generalized DBI field with $\phi$. We fix the initial values
of $\dot{\phi}$, $\phi$ and $\rho_m/H^2$. By this way, the initial
values of $Y$ and $Z$ are determined. Fig.~(2) shows that when the
initial speed $\dot{\phi_i}$ is small enough, the scalar would
acquire damped oscillating due to the Hubble friction in the false
vacuum ($\sigma_1$) and finally it is frozen in this vacuum. With
the increasing of initial speed, the scalar crosses the first
local maximum ($\xi_1$) and finally is frozen in the second false
vacuum ($\sigma_2$) (see Fig.~(3)). With even much larger initial
velocity, the scalar crosses the two local maximum ($\xi_1$ and
$\xi_2$) and finally dwells on the real vacuum ($\sigma_3$) (see
Fig.~(4)).

In Fig.~(5), we plot the evolution of the equation of state $w$ of
the generalized DBI field corresponding to Fig.~(4). We see the
generalized DBI field previously behaves as the dust matter and
then oscillating approaches $-1$ after a sufficient long time. The
reason for oscillating of $\omega$ can be understood as follows.
Eq.~(25) tells us when the speed of DBI field vanishes, the
equation of state is $-1$. Fig.~(4) shows there are many times for
the vanishing of speed during the damped oscillating. Every time
the generalized DBI field acquires vanishing velocity, its
equation of state is $-1$.

\begin{figure}[h]
\begin{center}
\includegraphics[width=9cm]{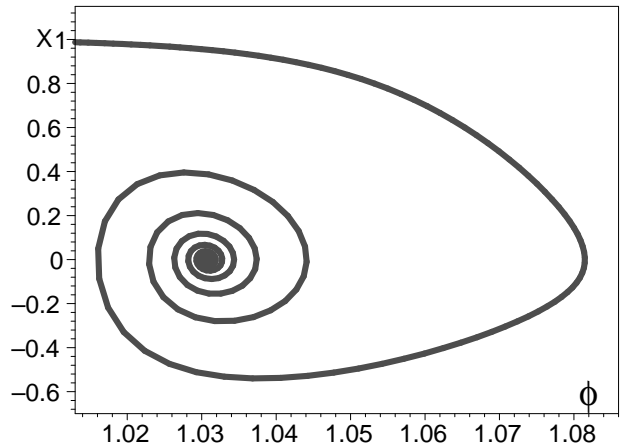}

\caption{Evolution of the speed ($X=\dot{\phi}$) of generalized
DBI field with $\phi$. The point $(\phi=\sigma_1=1.03,\ X=0)$ is a
stable spiral and thus an attractor. In this case, the generalized
DBI field behaves as a damping oscillator in the false vacuum
($\sigma_1$) and finally is frozen.}. \label{f2}
\end{center}
\end{figure}

\begin{figure}[h]
\begin{center}
\includegraphics[width=9cm]{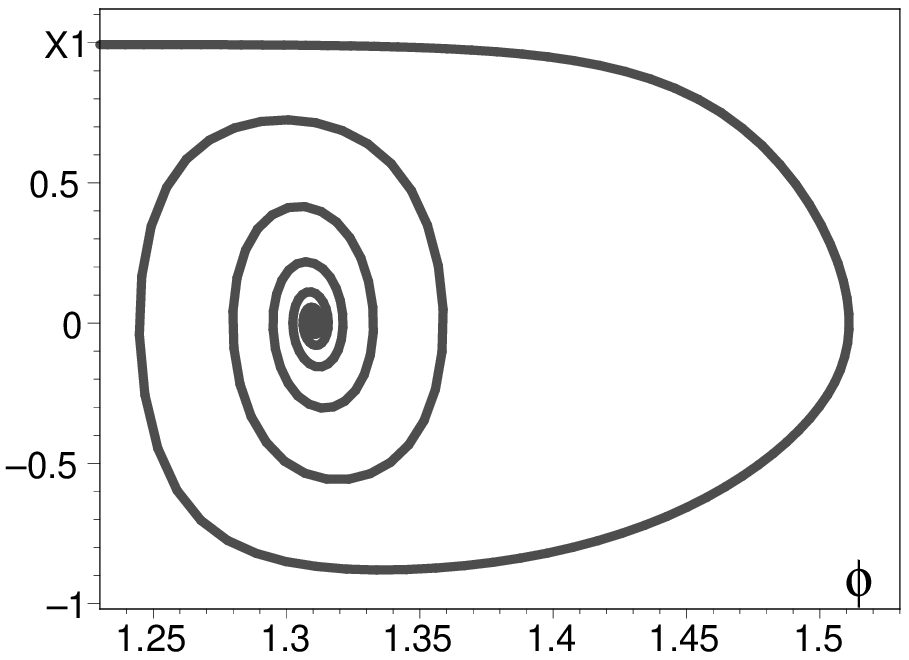}

\caption{Evolution of the speed ($X=\dot{\phi}$) of generalized
DBI field with $\phi$.  The point $(\phi=\sigma_2=1.31,\ X=0)$ is
a stable spiral and thus an attractor. In this case, the
generalized DBI field behaves as a damping oscillator in the false
vacuum ($\sigma_2$) and finally is frozen.}. \label{f3}
\end{center}
\end{figure}

\begin{figure}[h]
\begin{center}
\includegraphics[width=9cm]{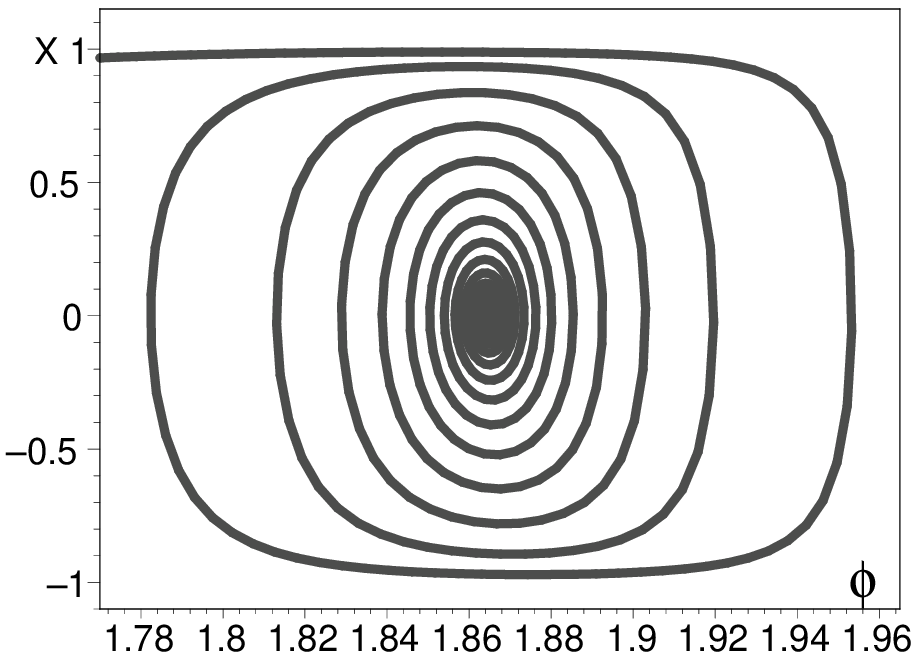}

\caption{Evolution of the speed ($X=\dot{\phi}$) of generalized
DBI field with $\phi$. The point $(\phi=\sigma_2=1.86,\ X=0)$ is a
stable spiral and thus an attractor. In this case, the generalized
DBI field behaves as a damping oscillator in the real vacuum
($\sigma_3$) and finally is frozen.}. \label{f4}
\end{center}
\end{figure}

\begin{figure}[h]
\begin{center}
\includegraphics[width=9cm]{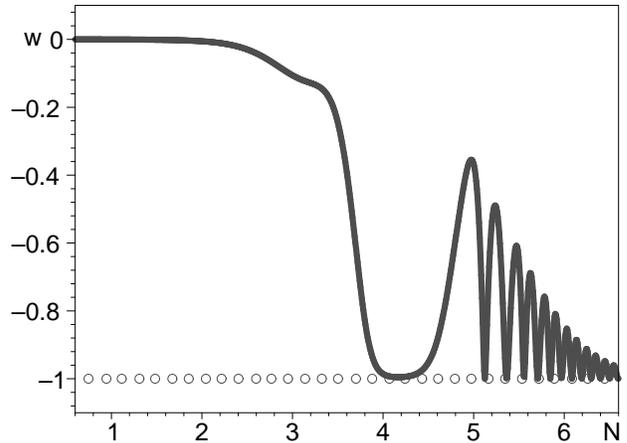}

\caption{The evolution of the equation of state for the
generalized DBI scalar field. It behaves as the dust matter at
higher redshifts and oscillating approaches $-1$ at the lower
redshifts}. \label{f5}
\end{center}
\end{figure}
\section{quintessence}

In this section, we shall present the method for dealing with
arbitrary quintessence potential. As an example, we would explore
the cosmic evolution of quintessence field with multiple vacua. To
this end, let's focus on the potential as follows
\begin{eqnarray}\label{quin-pot}
V=V_0
e^{b\left(\phi-a_1\right)\left(\phi-a_2\right)\left(\phi-a_3\right)
\left(\phi-a_4\right)\left(\phi-a_5\right)\left(\phi-a_6\right)}\;,
\end{eqnarray}
where $V_0,\ b,\ a_i$ are positive constants. As an example, we
consider, $V_0=1,\ b=1,\ a_1=1/2,\ a_2=3/5,\ a_3=1,\ a_4=17/10,\
a_5=2,\ a_6=5/2$.

\begin{figure}[h]
\begin{center}
\includegraphics[width=9cm]{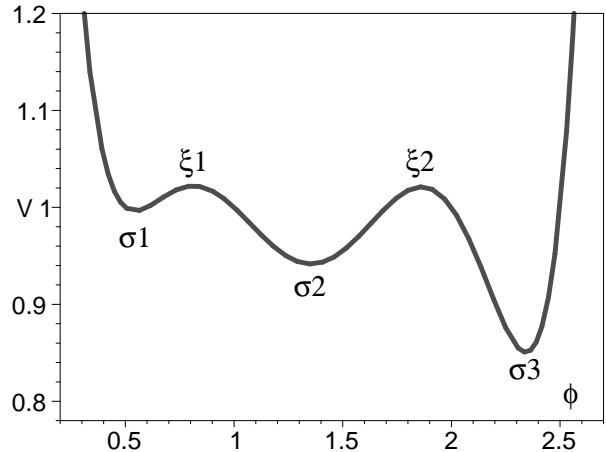}
\caption{The quintessence potential $V(\phi)$ with respect to
$\phi$. There are two local maximum ($\xi_1=0.813,\ \xi_2=1.862$),
one real vacuum ($\sigma_3=2.342$) and two false vacuum
($\sigma_1=0.545,\ \sigma_2=1.355$) for the potential. Physically,
the quintessence would firstly damped oscillates between these
vacua and finally dwell on one of the vacuum due to the Hubble
friction.}. \label{f6}
\end{center}
\end{figure}

In Fig.~{\ref{f6}}, we plot the potential $V(\phi)$ with respect
to $\phi$. There are two local maximum ($\xi_1=0.813,\
\xi_2=1.862$), one real vacuum ($\sigma_3=2.342$) and two false
vacuum ($\sigma_1=0.545,\ \sigma_2=1.355$) for the potential.
Physically, the scalar field would roll down the potential and
then damped oscillates between these vacua. Given an initial
finite velocity $\dot{\phi}$ and field value $\phi_0$, the fate of
the scalar is expected to dwell on the one of the vacuum due to
the Hubble friction. In what follows, we shall show theses points
numerically.

The Einstein equations and the equation of motion of the
quintessence are given by
\begin{eqnarray}\label{eq:quin00}
&&3H^2=\kappa^2\left(\frac{1}{2}\dot{\phi}^2+V+\rho_m\right)\;,\nonumber\\
&&2\dot{H}+3H^2=-\kappa^2\left(\frac{1}{2}\dot{\phi}^2-V+\omega_m\rho_m\right)\;,
\end{eqnarray}
and
\begin{eqnarray}\label{eq:quin11}
\ddot{\phi}+3H\dot{\phi}+V_{,\phi}=0\;,
\end{eqnarray}
respectively. Here $\rho_m$ and $\omega_{m}$ are the energy
density and the equation of state for the matter sources. In this
section, we also consider the case of dust matter, $\omega_m=0$.

Observing Eqs.~(\ref{eq:quin00}) and Eq.~(\ref{eq:quin11}), we
could absorb the constant $\kappa^2/3$ ($\kappa^2=8\pi$) into
$\phi^2, V$ and $\rho_m$, respectively,

\begin{eqnarray}
&&\phi^2\longrightarrow \phi^2\cdot\frac{6}{\kappa^2}\;,\\
&&V\longrightarrow V\cdot\frac{3}{\kappa^2}\;,\\
&&\rho_m\longrightarrow \rho_m\cdot\frac{3}{\kappa^2}\;.
\end{eqnarray}
Then above equations of motion turns out to be
\begin{eqnarray}\label{eq:ein0}
&&H^2=\dot{\phi}^2+V+\rho_m\;,\\
&&2\dot{H}/3+H^2=-\dot{\phi}^2+V-\omega_m\rho_m\;,\\
&&\ddot{\phi}+3H\dot{\phi}+\frac{1}{2}{V_{,\phi}}=0\;.
\end{eqnarray}
Given the scalar potential $V(\phi)$ and the equation of state
$\omega_m$, we are left with three variables, $a(t),\rho_m$ and
the quintessence $\phi(t)$. Then we have three variables and three
independent differential equations. So the system of equations is
closed.

The same as the DBI case, it is rather involved to find the
analytic solutions to the equations of motion (32-34). Hence in
order to solve them numerically, we had better rewrite them in the
autonomous form. To our knowledge, one have only explored some
special form of the quintessence potential, namely, the
exponential potential \cite{cope:98,ng:01}, power-law type
potential \cite{ratra:88,cald:98}, the Albrecht and Skordis
potential \cite{albrecht:99} an so on. For \emph{arbitrary}
potential, one have not yet find a general method to write the
equations of motion in the autonomous form. In what follows, we
shall propose a method that can be used to deal with
\emph{arbitrary} potentials.

To this end, we introduce the following dimensionless quantities
\begin{eqnarray}
&&X\equiv\frac{\dot{\phi}}{H}\;,\ \ \ \ \ \ \ \ Q\equiv\frac{{V_{,\phi}}}{V}=Q\left(Z\right)\;,\\
&&Y\equiv\frac{\sqrt{V}}{H}\;,\ \ \ \ \ \ N\equiv\ln a\;,\\
&&Z\equiv\phi\;.\
\end{eqnarray}

With the aid of these definitions, we can write the equations of
motion in the autonomous form with \emph{arbitrary} potentials:
\begin{eqnarray}\label{eq:quinaut}
\frac{dX}{dN}&=&-3X-Y^2Q-X\cdot\frac{\dot{H}}{H^2}\;,\nonumber\\
\frac{dY}{dN}&=&\frac{1}{2}XYQ-Y\frac{\dot{H}}{H^2}\;,\nonumber\\
\frac{dZ}{dN}&=&X\;,
\end{eqnarray}
with
\begin{eqnarray}
\frac{\dot{H}}{H^2}&=&-3X^2-\frac{3}{2}\left(1+\omega_m\right)\left(1-X^2-Y^2\right)\;.
\end{eqnarray}
We note that $Q$ is the function of $Z$. Therefore,
Eqs.~(\ref{eq:quinaut}) is indeed an autonomous system of
equations. In Table II, we present the properties of the five
fixed points for the quintessence. The points (a,\ c,\ e)
correspond to the false vacua ($\sigma_1,\ \sigma_2$) and real
vacuum ($\sigma_3$). The three points are stable spirals.  On
these epoches, the quintessence behaves as a damping oscillator
with the equation of state of firstly behaving as the dust matter
and then oscillating approaching $-1$. The points (b,\ d,)
correspond to the two local maximum ($\xi_1,\ \xi_2$) and they are
saddle points. On these points, the speed of the scalar field
exactly vanishes and the quintessence acquires the equation of
state of cosmological constant.
\begin{table*}[t]
\begin{center}
\begin{tabular}{|c|c|c|c|c|c|c|c|c|}
\hline Name &  $X$ & $Y$ &$Z$ & $\phi$
 & Stability & $\Omega_\phi$
 & $w_\phi$  \\
\hline \hline (a) & $0$ & $1$ & $0.545$ & $\sigma_1=0.545$ &
Stable spiral (attractor)
& $1$ & $-1$\\
\hline \hline (b) & $0$ & $1$ & $0.813$ & $\xi_1=0.813$ & Saddle
point
& $1$ & $-1$\\
\hline \hline (c) & $0$ & $1$ & $1.355$ & $\sigma_2=1.355$ &
Stable spiral (attractor)
& $1$ & $-1$\\
\hline \hline (d) & $0$ & $1$ & $1.862$ & $\xi_2=1.862$ & Saddle
point
& $1$ & $-1$\\
\hline \hline (e) & $0$ & $1$ & $2.342$ & $\sigma_3=2.342$ &
Stable spiral (attractor)
& $1$ & $-1$\\
\hline
\end{tabular}
\end{center}
\caption[crit]{Properties of the critical points for the
  quintessence potential given by
  Eq.~(26).} \label{crit0}
\end{table*}

In Fig.~(7-9), we plot the evolution of the rescaled speed, $X$ of
the quintessence with $\phi$. The figures show that, with the
increasing of initial speed, the quintessence is frozen in the
first false vacuum, the second false vacuum and the real vacuum,
respectively. Compared to evolution of DBI field,  we see from
Figs.(2-4) that the DBI field has more times of oscillations than
quintessence. How to understand this point? The equations of
motion tell us the friction term due to Hubble expansion is
$3H\dot{\phi}$ and $3H\dot{\phi}\left(1-\dot{\phi}^2\right)$ for
quintessence and DBI field, respectively. Then with the increasing
of speed $\dot{\phi}$, the friction term of DBI field is greatly
decreased. So the DBI field acquires more times of oscillations.

\begin{figure}[h]
\begin{center}
\includegraphics[width=9cm]{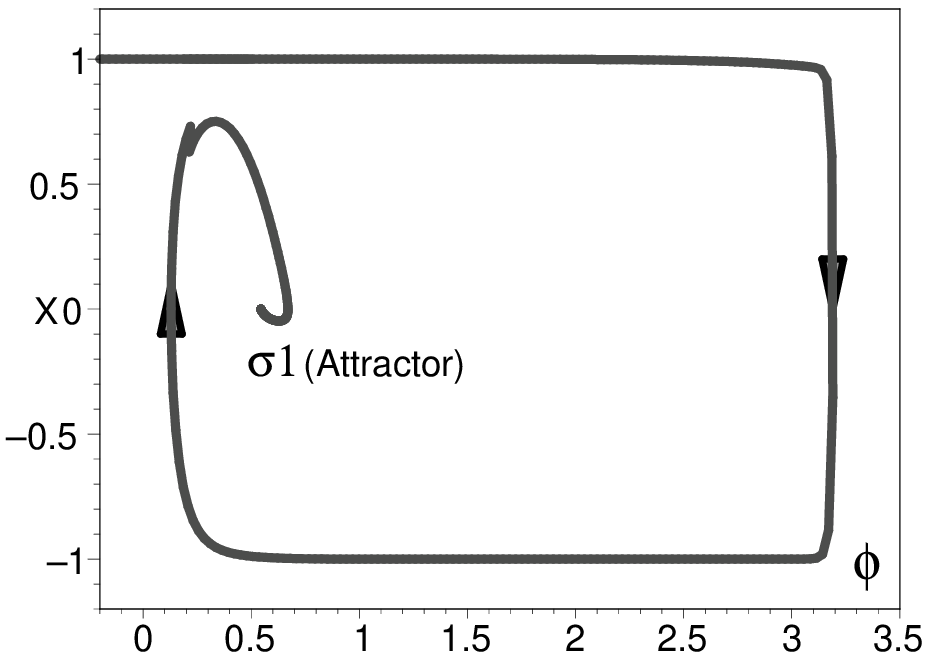}

\caption{Evolution of the rescaled speed ($X$) of quintessence
with $\phi$. The point $(\phi=\sigma_1,\ X=0)$ is a stable spiral
and thus an attractor. The quintessence is finally frozen at
$\phi=\sigma_1$.}. \label{f7}
\end{center}
\end{figure}

\begin{figure}[h]
\begin{center}
\includegraphics[width=9cm]{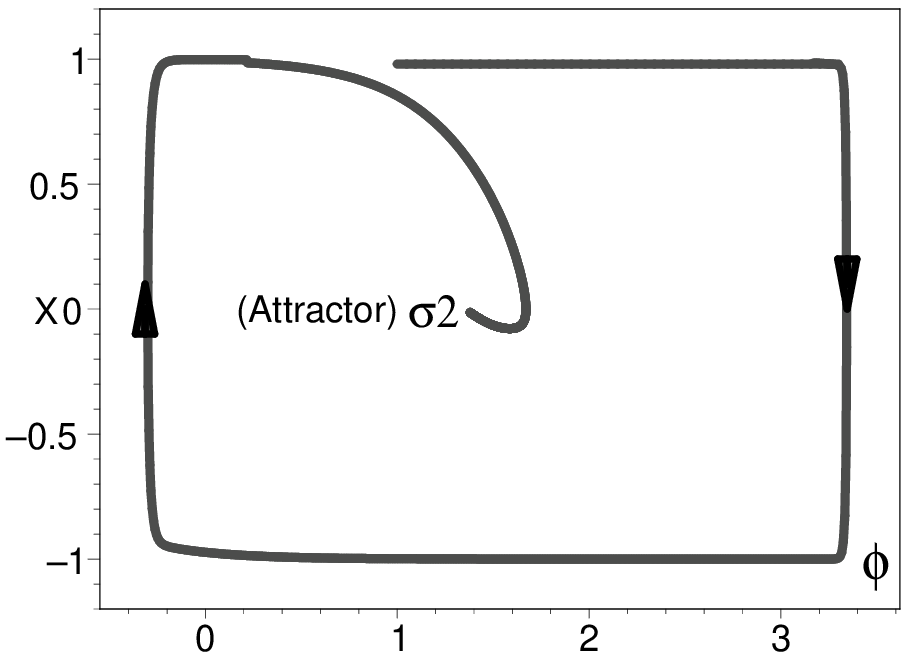}

\caption{Evolution of the rescaled speed ($X$) of quintessence
with $\phi$. The point $(\phi=\sigma_2,\ X=0)$ is a stable spiral
and thus an attractor. The quintessence is finally frozen at
$\phi=\sigma_2$.}. \label{f8}
\end{center}
\end{figure}

\begin{figure}[h]
\begin{center}
\includegraphics[width=9cm]{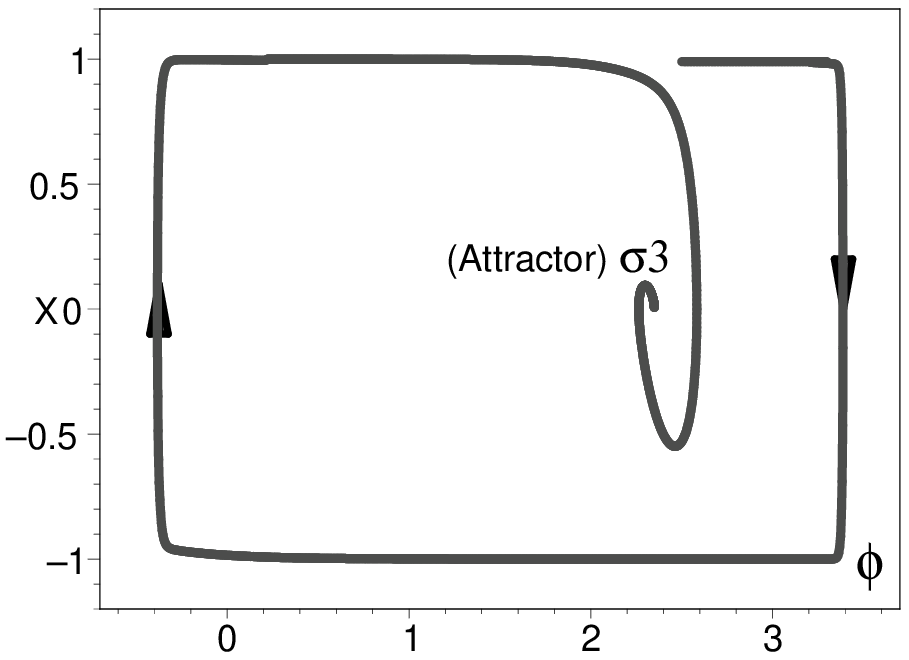}

\caption{Evolution of the rescaled speed ($X$) of quintessence
with $\phi$. The point $(\phi=\sigma_3,\ X=0)$ is a stable spiral
and thus an attractor. The quintessence is finally frozen at
$\phi=\sigma_3$.}. \label{f9}
\end{center}
\end{figure}
\section{conclusion and discussion}

In general, the equations of motion for generalized DBI field and
quintessence are rather complicated. So one resort to the method
of phase analysis by writing the equations of motion in the
autonomous form. Many special form of potentials have been studied
for DBI field \cite {guo:08} and quintessence
\cite{cope:98,cald:98,zla:99,mac:00,ng:01,cor:03,cald:05,linder:06,
bar:00,cope:09,liddle:99,sahni:00IJ,sahni:00PRD}. However, the
general method for dealing with \emph{arbitrary} potentials have
not yet been proposed. Thus the outcome of this article is that we
have found the method.

Different from the potentials studied in
Refs.~\cite{cope:98,cald:98,zla:99,mac:00,ng:01,cor:03,cald:05,linder:06,
bar:00,cope:09,liddle:99,sahni:00IJ,sahni:00PRD,guo:08}, we
investigate the cosmic evolution of the generalized DBI field and
quintessence with the potential of multiple vacua. We find that,
with the increasing of initial speed, both generalized DBI field
and quintessence are successively frozen, in the first false
vacuum, the second false vacuum and the real vacuum, respectively.
Compared to the evolution of quintessence, the generalized DBI
field has more times of oscillations. The reason for this point is
that, with the increasing of speed $\dot{\phi}$, the friction term
of generalized DBI field is greatly decreased. Thus the
generalized DBI field acquires more times of oscillations.

The conclusion in this paper may be trivial, but the proof and the
method are not. As an example, one could study the Coleman-De
Luccia tunnelling using this method. After making the Wick
rotation, $t=-i\tau$ in Eqs.~(21) and Eqs.~(38), we are able to
study the Coleman-De Luccia tunnelling numerically but exactly.

\acknowledgments

We thank one of the referees for pointing out some important typos.  This work is supported by the Chinese MoST 863 program under grant
2012AA121701, the NSFC under grant 11373030, 10973014, 11373020
and 11465012.

\newcommand\ARNPS[3]{~Ann. Rev. Nucl. Part. Sci.{\bf ~#1}, #2~ (#3)}
\newcommand\ASS[3]{~Astrophys. Space. Sci.{\bf ~#1}, #2~ (#3)}
\newcommand\AL[3]{~Astron. Lett.{\bf ~#1}, #2~ (#3)}
\newcommand\AP[3]{~Astropart. Phys.{\bf ~#1}, #2~ (#3)}
\newcommand\AJ[3]{~Astron. J.{\bf ~#1}, #2~(#3)}
\newcommand\APJ[3]{~Astrophys. J.{\bf ~#1}, #2~ (#3)}
\newcommand\APJL[3]{~Astrophys. J. Lett. {\bf ~#1}, L#2~(#3)}
\newcommand\APJS[3]{~Astrophys. J. Suppl. Ser.{\bf ~#1}, #2~(#3)}
\newcommand\JHEP[3]{~JHEP.{\bf ~#1}, #2~(#3)}
\newcommand\JCAP[3]{~JCAP. {\bf ~#1}, #2~ (#3)}
\newcommand\LRR[3]{~Living Rev. Relativity. {\bf ~#1}, #2~ (#3)}
\newcommand\MNRAS[3]{~Mon. Not. R. Astron. Soc.{\bf ~#1}, #2~(#3)}
\newcommand\MNRASL[3]{~Mon. Not. R. Astron. Soc.{\bf ~#1}, L#2~(#3)}
\newcommand\NPB[3]{~Nucl. Phys. B{\bf ~#1}, #2~(#3)}
\newcommand\CQG[3]{~Class. Quant. Grav.{\bf ~#1}, #2~(#3)}
\newcommand\PLB[3]{~Phys. Lett. B{\bf ~#1}, #2~(#3)}
\newcommand\PRL[3]{~Phys. Rev. Lett.{\bf ~#1}, #2~(#3)}
\newcommand\PR[3]{~Phys. Rep.{\bf ~#1}, #2~(#3)}
\newcommand\PRD[3]{~Phys. Rev. D{\bf ~#1}, #2~(#3)}
\newcommand\RMP[3]{~Rev. Mod. Phys.{\bf ~#1}, #2~(#3)}
\newcommand\SJNP[3]{~Sov. J. Nucl. Phys.{\bf ~#1}, #2~(#3)}
\newcommand\ZPC[3]{~Z. Phys. C{\bf ~#1}, #2~(#3)}
\newcommand\IJMPD[3]{~Int. J. Mod. Mod. Phys. D{\bf ~#1}, #2~(#3)}
 \newcommand\IJGMP[3]{~Int. J. Geom. Meth. Mod. Phys.{\bf ~#1}, #2~(#3)}
  \newcommand\GRG[3]{~Gen. Rel. Grav.{\bf ~#1}, #2~(#3)}

\end{document}